\documentstyle[12pt]{article}
\begin{document}

\newcommand{\be}{\begin{equation}}
\newcommand{\ee}{\end{equation}}
\newcommand{\ba}{\begin{eqnarray}}
\newcommand{\ea}{\end{eqnarray}}
\newcommand{\p}{\partial}
\newcommand{\f}{\frac}
\newcommand{\s}{\sqrt}

\begin{center}
{\bf The Graviton Production in a Hot Homogeneous Isotropic
Universe}\\ 
\vskip \baselineskip
{\bf Gerstein S.S., Logunov A.A., Mestvirishvili M.A.}\\

\vspace*{1mm}
{\it Division of Theoretical Physics, IHEP, 
Protvino, Russia}\\
\end{center}

\vspace*{5mm}
\begin{abstract}
It is shown that the RTG predicts an opportunity of the intensive
production of gravitons at the early stage of evolution of the
homogeneous isotropic Universe. A hypothesis is suggested that the
produced gas of gravitons could be just the ``dark matter'' which
presently manifests itself as a ``missing mass'' in our Universe.
\end{abstract}

\vspace*{10mm}

In publications~[1,2]  an opportunity  of graviton production in the
Universe  was studied in detail. In paper~[2]  the following formula
for the rate of graviton production in the homogeneous and isotropic
Universe was derived
\be \f{1}{\sqrt{-g}}\;\f{d}{dt}(\s{-g}n_g)=\f{1}{288\pi}R^2, \ee
where it was supposed that
\be \f{R^2}{R_{\rho\lambda\mu\nu}R^{\rho\lambda\mu\nu}}\ll 1, \ee
and $R $ is the scalar curvature,  $R _ {\rho\lambda\mu\nu} $ is the
Riemannian curvature tensor.

The following equation of state takes place in the hot Universe for
radiation dominated stage of its evolution 
\be p=\f{1}{3}\rho c^2. \ee

But as, according to the  General Relativity  Theory (GRT), the
scalar curvature $R $ is exactly zero at this stage of the Universe
evolution, the authors of papers~[1,2] came to the conclusion that
production of gravitons in the hot homogeneous and isotropic Universe
does not occur. In publication~[1] attention was paid also to the
fact that the production of gravitons apparently forbids isotropic
singularities, in the vicinity of which the equation of state should
be as follows
\be p>\f{1}{3}\rho c^2. \ee

This conclusion has apparently arisen because in this case the scalar
curvature $R $ would become arbitrary large, and therefore there
should be an extremely intensive production of gravitons, and
consequently in presence of a singularity this would result in
contradiction to the modern data on the density of matter in the
Universe.

In the Relativistic Theory of Gravitation (RTG), which views a
gravitational field as a physical one with  spins 2 and 0 and
propagating in the Minkowski space, completely another situation
arises: the evolution of the homogeneous and isotropic Universe is
determined by other equations~[3,4] and (extremely important) here
there are no singularities:
\be \f{1}{a}\;\f{d^2a}{d\tau^2}=-\f{4\pi G}{3} \left (\rho +
\f{3p}{c^2}\right
) -2\omega \left (1- \f{1}{a^6}\right ), \ee \be H^2\equiv \left (
\f{1}{a}\;\f{da}{d\tau} \right )^2 =\f{8\pi G}{3}\rho
-\f{\omega}{a^6} \left
(1- \f{3a^4}{a^4_{\max}}+2a^6\right ), \ee
where
\be \omega =\f{1}{12} \left ( \f{mc^2}{\hbar} \right )^2, \;m\;
\mbox{is the
graviton mass}. \ee

It follows~[3,4] from these equations that for a radiation dominated
stage of the Universe evolution in the domain of the small values of
the scale factor $a (\tau) $ the following equation takes place:
\be \hspace*{-1cm} \f{\ddot a}{a} +\left ( \f{\dot a}{a} \right )^2=
\f{\omega}{a^6},\;\;\mbox{where}\;\; \dot a=\f{da}{d\tau}. \ee

In the GRT  l.h.s. of  Eq. (8) in radiation dominated region is
exactly zero, and therefore, Friedman stage  takes place, if  scale
factor $a (\tau) $ varies according to the law $ \sqrt {\tau} $. In
RTG, according to Eq.(8), there is an ``under-Friedman'' stage in the
radiation dominated phase  of evolution of the Universe. The scalar
curvature $R $ for the homogeneous and isotropic Universe is as
follows
\be R=-\f{6}{c^2} \left [ \f{\ddot a}{a}+ \left ( \f{\dot a}{a}\right
)^2\right
]. \ee

On the basis of Eq. (8) we have
\be R=-\f{1}{2} \left ( \f{mc}{\hbar} \right )^2 \f{1}{a^6}. \ee

From  equations (6) it follows that the scale factor $a (\tau) $
cannot become  zero, and its minimal value is as follows \be
a_{\min}= \left ( \f{\rho_{\min}}{2\rho_{\max}} \right )^{1/6}, \ee
where \be \rho_{\min} =\f{1}{16\pi G} \left ( \f{mc^2}{\hbar} \right
)^2, \ee and the maximal density of matter in the gravitational field
$ \rho _ {\max} $ is in fact an integral of  motion and it is not
determined by the theory.

On the basis of Eqs. (10), (11) and (12) it follows that at the
moment when the maximal density of matter is reached, the scalar
curvature of effective Riemannian space takes the following value
\be R=-\f{16\pi G}{c^2}\rho_{\max}. \ee

At this instant of time the Hubble ``constant'' $H $ is precisely
zero. We can see from  formula (13) that in the RTG, just opposite to
the  GRT, the scalar curvature $R $ in radiation dominated stage of
the Universe evolution  is not  zero. Moreover, it can be large
enough  as it is determined by the peak density of matter $ \rho _
{\max} $ in the gravitational field.

Thus, according to RTG, in the radiation dominated phase  of the
Universe evolution there is an ``under-Friedman'' stage, in which the
scalar curvature $R $ is not only different from zero, but also can
be large enough, as it is determined by the peak density of matter $
\rho _ {\max} $. To determine the rate of graviton production we
cannot take advantage of  formula (1), as it is obtained in
approximation (2), which in our case is not fulfilled.

If on the base of dimensional reasons we assume that the rate of
graviton production in general depends only on the following
quantities 
\be R^2,\; R_{\rho\lambda\mu\nu} R^{\rho\lambda\mu\nu}, \ee 
than it is necessary to pick such a time interval, during which the
Hubble ``constant'' reaches the  maximum, as after that there soon
occurs the Friedman stage. From  Eqs. (5) and (6) it is easy to find,
that $H $ reaches its maximum at the instant of time  when the scale
factor $a (\tau) $ is as follows
\be a^2(\tau) =\f{3}{2}a^2_{\min}. \ee
By using Eq. (15) from Eqs. (6) it is discovered that the peak value
of the Hubble ``constant'' is as follows
\be H_{\max} =3^{-2} (32\pi G\rho_{\max})^{1/2}. \ee

At the instant of time when $H $ reaches the maximum the scalar
curvature $R $ is as follows 
\be R=-\left (\f{2}{3}\right )^316\pi G\f{\rho_{\max}}{c^2}, \ee
$ \f {\ddot a} {a} $ is determined by expression
\be \left ( \f{\ddot a}{a}\right ) =3^{-4}\cdot 32 \pi G \rho_{\max}.
\ee

The invariant obtained by convolution of a curvature  tensor
determined from  metric of the homogeneous and isotropic Riemannian
space is as follows
\be R_{\rho \lambda\mu\nu} R^{\rho\lambda\mu\nu} =\f{12}{c^{4}} \left
[ \left (
\f{\ddot a}{a} \right )^2 + \left ( \f{\dot a}{a} \right )^4 \right
]. \ee 
By substituting  (16) and (18) into this expression we obtain
 \be R_{\rho \lambda\mu\nu} R^{\rho\lambda\mu\nu} =8\cdot 3^{-7}
 \left (
\f{32\pi G}{c^2}\rho_{\max} \right )^2. \ee

It is necessary to mention that the Hubble ``constant'' varies from
zero value to the peak value $H _ {\max}$  determined by  formula
(16) for a rather small time interval given by the following
expression~[3,4]:
\be \tau = \left ( \f{3}{32\pi G\rho_{\max}} \right )^{1/2} \cdot
\left [
\f{\s3}{2} +\ln \left ( \s{\f{3}{2}}+\s{\f{1}{2}} \right ) \right ].
\ee

If the rate of graviton production is determined by quantities (14)
there can be created enough large number of gravitons for a time
interval (21), if density $ \rho _ {\max} $ will have a significant
value. But if it will be much less than the Planck density then
produced gravitons at once become free, and their energy further will
diminish due to the red shift.

Thus, there should arise a relativistic relict gravitational
background of a non-thermal origin. The gravitons interact among
themselves strongly enough as their coupling constant is equal to
unity. This circumstance can break homogeneity of a relict
gravitational background of a non-thermal origin at sufficient
density of gravitons. From dimensional reasons the general number  of
produced quanta of the gravitational field in cubic centimeter of
volume will be proportional to the following quantities \be
cR^2\tau,\;\; c(R_{\rho\lambda\mu\nu} R^{\rho\lambda\mu\nu})\tau, \ee
where  $R^2, R _ {\rho\lambda\mu\nu} R ^ {\rho\lambda\mu\nu}, \tau $
are given by expressions (17), (20) and (21). It follows from these
formulas  that the rate of  graviton production in the hot radiation
dominated phase of evolution of the Universe is mainly determined by
the peak density of matter $ \rho _ {\max} $. Knowing more detailed
pattern of  graviton production  it would be possible to spot the
peak value of density of matter, which the Universe  had in the
present cycle of ``expansion''. On the other hand, it is possible to
state a hypothesis: the gravitational background of a non-thermal
origin could be just that ``dark matter'', which manifest itself as
``missing mass'' in our Universe . But all this requires more careful
analysis.

The authors are grateful to A.A. Grib, V.A. Petrov, N.E. Tyurin and
Yu.V. Chugreev for valuable discussions.

\end{document}